\documentclass[amsmath,amssymb,aps,pra,reprint,groupedaddress,showpacs]{revtex4-1}

\usepackage[english,german]{babel}
\usepackage[latin1]{inputenc}
\usepackage[T1]{fontenc}

\usepackage{multirow}
\usepackage{verbatim}
\usepackage{color,graphicx}
\usepackage{xcolor}

\usepackage[printwatermark]{xwatermark}
\newwatermark[allpages,color=black!20,angle=45,scale=2,xpos=0,ypos=0]{Draft - to be submitted}

\begin{document}

\title{Filling missing data in point clouds by merging structured and unstructured point clouds}
\email{This article is to be submitted soon. Further application of the theory is to be published soon }
\author{Franziska Lippoldt}
\email[]{lippoldt331@gmail.com}
\author{Hartmut Schwandt}
\email{schwandt@math.tu-berlin.de}

\selectlanguage{german}
\affiliation{Fachbereich Mathematik, Technische Universität Berlin, D-10623 Berlin, Germany}
\selectlanguage{english}
\date{\today}

\begin{abstract}
Point clouds arising from structured data, mainly as a result of CT scans, provides special properties on the distribution of points and the distances between those. Yet often, the amount of data provided can not compare to unstructured point clouds, i.e. data that arises from 3D light scans or laser scans. This article hereby proposes an approach to extend structured data and enhance the quality by inserting selected points from an unstructured point cloud. The resulting point cloud still has a partial structure that is called "half-structure". In this way, missing data that can not be optimally recovered through other surface reconstruction methods can be completed.
\end{abstract}

\maketitle

\section{Introduction}
Recent development of technology has opened up wide possibilities to create 3D data of a real world object. Medical imaging and CT scanning was one of the first technologies to construct 3D data of the human body. Strongly supported by the idea of photographic images, a CT scan consist of a staple of images that together form a box in 3D space.\\
Development of 3D scanning methods has always been application driven. Given a specific idea or need, a 3D scanner was developed to obtain data. This also explains the huge variety of possible 3D scanning technologies and systems nowadays. From light scanners and laser scanners to CT scanners, technology can be chosen that is most appropriate for the task.\\
Whatever of those 3D scanning technologies is chosen, several restrictions apply. The theoretical maximum resolution of scanning devices could be achieved in ideal circumstances. For light based 3D scanners, the environment plays an important role, including the structure and surface of the object as well as surrounding light. \\
The product of a 3D scan is either grid data or a point cloud, which then can be reconstructed to a surface. The movement of the scanner using light can influence the quality and exactness of the point cloud. Yet, repeated scanning of the same areas will not leaded to a better quality every time. The idea that more points in the point clouds lead to a better scan is not appropriate. In fact, a lot of points on the point cloud are not useful after the point alignment has been made: for photographic based scanners, typical density lines develop around each single shot of an image, leading to unnecessary stripes on the model.\\
A significant amount of research in computer vision has been done on surface reconstruction, few the point cloud itself turns into a more and more interesting object. Given the fact that the data should be printed, the creation of a surface is of high value. But this does not imply the necessity to improve the surface reconstruction algorithm. Most algorithms are developed to minimise scanner typical errors or unavoidable environmental influences on the resulting surface, but this article is proposing a method for preprocessing the point cloud itself. This is most useful when available tools are available, i.e. CT scanner and 3D light or laser scanner, and in fact is desirable for reaching a high quality point cloud and hence forth optimal reconstructed surface.\\
The quality and the number of points in a point cloud are not automatically correlated. 
This article will explore the idea of inducing the structure on the point cloud by using CT scan data as a reference. The resulting point cloud will be called "half structured": typical scanning lines are erased, yet local density may vary according to the desired quality based on surface smoothness criteria.\\
There may be a huge choice of reconstruction methods available for specific problems and errors, but the final goal in the reconstruction of the surface is a general reliability. Not less was my work inspired by the fact that only few algorithms have actually been tested on large data sets. This half structured point cloud can be created for arbitrary large point clouds, its complexity mainly depends nearest neighbor search and geometric calculations using multi threads.\\
This paper will discuss the structure of point clouds. Those point clouds are supposed to be properly oriented, i.e. come not only with 3D coordinates but also with point normals. In fact, this article shows that the optimal way of substituting missing data in unstructured point clouds is the merging of a structured and an unstructured point cloud into a "half-structured" point cloud. Further common tasks such as downsampling and feature-based downsampling can be integrated into the creation of the "half-structured" point cloud and propose interesting features independent of the surface reconstruction.

\section{Related work}
So far, there exists research on missing data in reconstructed surfaces, downsampling of point clouds and specific research on the improvement of data from the specific light scanners or CT (see for example \cite{takeda1982fourier}). This paper proposes an approach that extends previous ideas. Filling missing data by point set structuring \cite{lafarge2013surface} is a feasible approach given that the point cloud already all necessary edges and only lacks data on relatively even parts. The same applies for the idea of edge-aware point set resampling as described by Huang et al \cite{huang2013edge}. \\
For research on surfaces that are hard to model due to rough structures, Kim \cite{kim2004modeling} has been developing a iterative level-set approach. In contrast, this paper proposes a direct method for solving the problem of roughness by additionally providing a second data set. Even though this article requires less data, the level-set approach is highly involved mathematical and requires the user to determine the number of steps for the main iteration. Convergence to an optimal model is not given, in fact it does not converge to a desired surface. One should also mention that the computation is complex and requires to insert the point cloud into a grid, i.e. turn it into structured data. \\
The fact that the reconstructed surface is expected to contain no cracks or holes has been so far discussed in the sense of optimizing surface reconstruction, see for example the proceedings of Kahzdan \cite{kazhdan2005reconstruction}, which later lead to the Poisson surface reconstruction method \cite{kazhdan2013screened}.\\
This article combines the ideas of finding an ideal crack-free surface and structuring the point cloud. In fact, it provides a basis to erase those cracks before reconstructing the surface, instead of regarding the  elimination  of holes as a surface reconstruction feature.

\begin{figure*}[ht] \label{fig:missingdata}
\centering
\includegraphics[width=180mm]{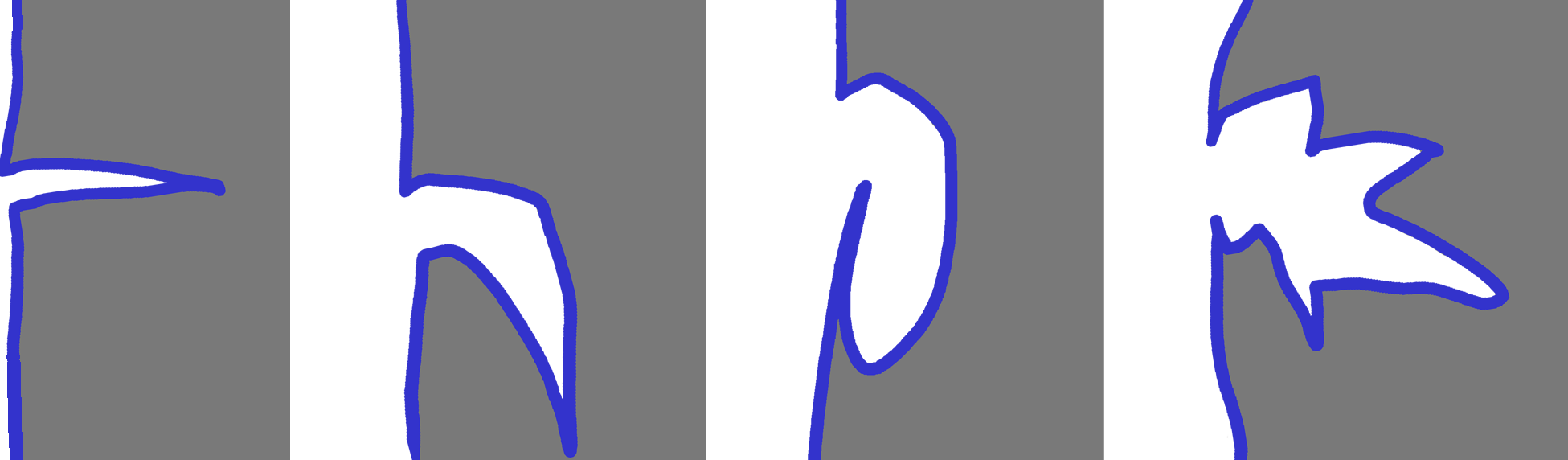}
\caption{Different cracks on an object, varying in level of complexity from left to right, except for the left example all the others will lead to a hole in the point cloud from structured light scans}
\end{figure*}

\section{Point cloud features and structure}
Given a surface, the basic notations of normals, curvature, holes has been clearly defined. This paper proposes an exact way for removing holes that would arise in the reconstructed surface, naturally without reconstructing the surface and independent of surface reconstruction methods.\\
There are several point cloud properties required for providing optimal results:
The point clouds need to be oriented, i.e. regard point cloud P a set of  N points of dimension 6: $P = \{ p_1, \cdots, p_N\} $, where each point contains its coordinates and normals $p_i = ( p_{coord}, p_{normal})$. For point clouds arising from light scanning technologies, normals are naturally calculated during the scanning process. \\
An important feature of point clouds is the arrangement of points and the distance between them. 
This article considers a point cloud as structured if it was reconstructed from grid data. Otherwise, it will be called unstructured.

\section{Global and local missing data}\label{sec:missingdata}
In this paper, we will regard any point cloud as a set of  N points $P = \{ p_1, \cdots, p_N\} $. This is a general definiton which both fits structured and unstructured point clouds. Further subsections will shortly illustrate why the finding and definition of "missing data" is not trivial.\\
Typically, "missing data" is only detected after the reconstruction of the surface. Given a reconstructured surface $S_R$, define missing data in $P$ as an error in surface properties of $S_R$. Those properties include the surface coordinates, the curvature and its normals. Those errors of $S_R$ may vary according to the surface reconstruction method. 
 
\subsection{Global missing data}
For structured point clouds, the critical aspect is the resolution of the according reconstructed surface.
An example for structured point clouds is the iso surface of a CT scan. The limit of the resolution depends mainly on two factors: the limitation of radiation during the scanning process, which limits the overall resolution of the scan and the effective implementation of the reconstruction via Marching cubes. For the latter problem, a multi-resolutional Marching Cubes version by Weber et al \citep{weber2003extraction} has been developed that allows local grid refinement. However, the limit of radiation still remains as the main factor for low resolution. \\
The additional insertion of points into a point cloud is also called up-sampling. The aim of up-sampling is not to combine those points through a straight line, but rather to estimate the curvature at a local area based on comparing points and normals.\\
Yet, an average iso-surface created from dicom data does not yield a "smooth" surface. This is a result from the basic idea of the marching cubes algorithm used to extract the surface, see \cite{lorensen1987marching}.
Instead, the surface contains small edges at areas that are slightly curved, comparable to a photo which   has a low resolution and is pixelated. This "pixelation" is a direct result of the structure of the marching cubes and the dicom data structure.

\subsection{Local missing data}
Recent 3D scanning methods using cameras or lasers allow to construct point clouds with high resolution. However, light based approaches will, even though they may be further developed, only scan object areas that are reachable by light. For complex objects, data might get lost on local areas that are complex, such as cracks or holes. Recent research has proposed various methods for surface reconstruction that estimate those wholes and creates an "optimal" surface. Yet, complex structures completely omitted can not be recovered as such.\\
For the purpose of better understanding the reason of local missing data, figure 1 illustrates examples.The left example shows an optimal surface crack, i.e. a crack that might be completely scanned. Even in case it is not, observing points and normals around the area of the crack are expected to lead to a authentic reconstruction of the surface. Determining the edge of the crack might be tricky and it could be smoothened to a certain degree, but specifying parameters on the smoothness of the crack will lead to satisfying results.\\
This is not the case for the other three examples. In example 2 and 3 counting from the left, the lower part of the surface cannot be scanned by any light scanning method. In fact, the only parts of the crack that can be scanned are the upper parts of the crack, i.e. everything a straight line can reach without touching the front part of the object. In fact, it is hard to determine whether the missing part is concave, convex or even more complex, as displayed in the right example.\\
Knowing that the part of the surface is either straight, concave or convex, one additional point indicating the curvature will be enough to reconstruct an approximate gap filling. \\
Example 4 in figure 1 shows that one crack in the object may actually induce several areas of local missing data. Again, for each area, one may not deliberately assume the curvature of the surface. Those lines displayed as straight may as well be concave or convex or a combination of both.

\section{Geometric approach features}\label{sec:algo}
The geometric approach for creating a "half-structured" point cloud is based on the missing data observations made in the previous chapter. Assuming that both point clouds do not only contain the 3D coordinates but also the estimated normals obtained during the cloud creation process, the point clouds can be locally treated as estimates for the resulting surface. \\
Let $P_{str}$ be the structured point cloud, let $P_{un}$ be the unstructured point cloud. Accordingly denote the points $p \in P_{str}$ , $q \in P_{un}$ accordingly, and $n_p, n_q$ their corresponding normals. Then the "half-structured" point cloud can be achieved by finding the optimal points
by checking for geometric conditions on the points $p,q$ and their normals $n_p,n_q$.\\
The nearest neighbor search is an essential part. It may be improved by using the approximate nearest neighbor search. Restrictions on the speed mainly depend on the efficiency of finding nearest neighbors. The normal criteria can be defined globally, but need to be adapted to the surface. The shape of the neighborhood influences the efficiency of the method. \\
The crucial detail is to find a point cloud that represents one surface. Given the fact that those two point clouds arise from the same object, one may assume that this is trivial. But actually, due to the different technological approach of both point clouds, their according reconstructed surfaces vary in volume i.e. scaling on all axes.\\
The following features can be implemented with geometric means:
\subsection{ Global downsampling} 
The evolving point cloud has two extremes: by setting the geometric criteria to an extreme, one may in the worst case setting obtain the structured or unstructured point cloud itself. Parameters need to be selected properly to create an appropriate cloud. Any point cloud that arises from the merging of the structured with the unstructured point cloud will have less or as many points as the unstructured point cloud. Global settings on the k-nearest neighbor search lead to a point cloud of at most $k*N$ points, where N is the number of points in the structured point cloud.
\subsection{ Feature based downsampling} 
The globally defined constraints on the distance and normal, i.e. tangent plane, lead to a local selection of the number of points. At extremely steep or nearly even areas, a higher density of points can be evoked. 
\subsection{ Detection of Outliers} 
Given that there are two point clouds as a reference for one object, a simple check of whether or not both clouds contain outliers can be performed. There are two possible settings for detecting outliers:
First, if the structured point cloud is reliable, i.e. contains nearly no errors, such as outliers or noise. Then the unstructured point cloud does not need to meet a lot of conditions regarding its errors, and the "half-structured" point cloud is purely based on the model of the structured point cloud, filtering the unstructured point cloud.\\
Second, the structured point cloud may contain noise or outliers. In this case the merging of the point clouds can also impose one condition on the structured point cloud. Given that the points of the structured point cloud are more sparse distributed then the other point cloud, the finding of outliers needs different parameters. This is also the case for undesired areas. If the structured point cloud arise from a CT scan, than scattered objects on the inside can be omitted using this criteria with appropriate parameters.Yet a reasonable clean CT scan can be created by choosing the right parameters during the scanning process. 
\subsection{ Complexity}
The overall calculation used only requires constant geometric calculations and finding of the nearest neighbors. The k-nearest neighbor search has average complexity $O(N \log N )$, where N is the number of points in the unstructured point cloud. For reference see the research published by Friedman \cite{friedman1977algorithm} and Bentley \cite{bentley1975multidimensional}. The geometric approach  applied after the nearest neighbor search has constant complexity $O(M k)$ with $M$ being the number of points of the structured point cloud. Worst case behavior for k-d tree search is only reached when the distance between the points is unreasonable irregular. For point clouds constructed by light scanning methods, a reasonable regular scanning process constructs reasonable point clouds and so the average complexity is expected to be achieved.

\section{ Maximal bounds of the point cloud}
\subsection{Maximal distance between nearest neighbors}
This section shows why the maximal distance between nearest neighbors of the resulting point cloud is bounded. Remark that the property of nearest neighbor is not reflexive, if point A is nearest to point B, point B does not necessarily be nearest to point A.\\
Let's start with the maximal distance of points from the structured point cloud. There exists a unique maximal distance $d_{struct}$ between nearest neighbors. The proof than shows that the nearest neighbors of the "half-structured" point cloud still have a maximal distance $d_{half}$. \\
This article assumed that structured point cloud arise from data lying on a grid. This especially means that for simply connected surfaces, the nearest neighbor for an arbitrary point is strictly bounded by a value $d_{struct}$, which is directly related to the size of the grid. For the standard marching cubes algorithm for example, the maximal distance between nearest neighbors is not larger than the length of the diagonal of the cube.\\
Now let $A,B$ be two points of the "half-structured" point cloud. Let B be the point closest to A. Let $d_{struct}$ be the maximal distance between nearest neighbors of the structured point cloud as denoted above. Let $d_{un}$ denoted the maximal distance between a structured cloud point  in $P_{struct}$ and the unstructured cloud point in $P_{un}$, used to define the neighborhood for selecting points, which is automatically given through the nearest neighbor search. Then for the maximal distance $d_{half}$ between $A$ and $B$ holds: \\
\[
d_{half}(A,B) \leq
\begin{cases}
 d_{struct} & \text{if A,B } \in P_{struct}\\
 d_{struct} + 2d_{un} & \text{if A,B}\in P_{un}\\
  d_{struct} + d_{un} & \text{else}\\
\end{cases}
\]
Then the distance $d_{half}$ between those points is bounded by the maximum of those three cases. 
For structured point clouds that include outliers, this approach does not hold in general. Given that the corresponding object is simply connected, every point that does not fulfill this condition can be specified as an outlier and extracted. 

\subsection{ Point cloud density}
Another special property of half-structured point clouds is the density, i.e. the maximal number of points within a certain neighborhood. For structured point clouds, that arise from grid data, a maximal number of points inside a local neighborhood, i.e. inside a cube defined by the grid, is given. Given the method as described above using k nearest neighbors, the maximal number of points within a cube as defined by the marching cubes algorithm with side length h yields a maximal number of 
\[
n_{max} = 12 \cdot k
\]
for a given cube c width width w, the maximal number of points contained is
\[
n_{max,c} =  12 \cdot  k \cdot {\lceil{\frac{w}{h} + 1} \rceil }^3 .
\]

\section{Conclusion}
The half structured point cloud combines the advantage of global maximal and local optimal resolution.
 Preparing the point cloud for critical missing data raises the quality of the surface to be reconstructed - independent of the surface reconstruction method used afterwards.\\
The features described in the previous chapter \ref{sec:algo} lead to the following result:
\begin{enumerate}
\item optimal and exact filling of missing data
\item noise or outlier reduction
\item control of the point cloud size in a reasonable way
\item maximal distance for two nearest neighbors
\end{enumerate}
all of which are independent of the surface reconstruction method chosen for further processing.
In fact, the reliable inclusion of missing data might lead to a reliable genus estimation of the surface, i.e. the number of "holes" in the original object. Regarding the fact that for a 'half-structured" point cloud, a maximal distance between the nearest two neighbors is given, the detection of holes in the object and hence point cloud is possible.

\bibliographystyle{siam}

\end{document}